\documentclass[sigconf]{acmart}
\AtBeginDocument{%
  }

\copyrightyear{2026}
\acmYear{2026}
\setcopyright{cc}
\setcctype{by}
\acmConference[CHI '26]{Proceedings of the 2026 CHI Conference on Human Factors in Computing Systems}{April 13--17, 2026}{Barcelona, Spain}
\acmBooktitle{Proceedings of the 2026 CHI Conference on Human Factors in Computing Systems (CHI '26), April 13--17, 2026, Barcelona, Spain}
\acmDOI{10.1145/3772318.3790676}
\acmISBN{979-8-4007-2278-3/2026/04}

\usepackage{booktabs}

\usepackage{soul}
\usepackage{enumitem}
\usepackage{hyperref}
\usepackage{xargs}      
\usepackage{multirow}
\usepackage{subcaption,siunitx,booktabs}
\usepackage{graphicx}
\usepackage[normalem]{ulem}
\useunder{\uline}{\ul}{}
\usepackage{lscape}
\usepackage{array}
\usepackage{lscape}
\usepackage{xcolor}

\definecolor{lightorange}{rgb}{1,0.8,0.4}
\definecolor{lightgreen}{RGB}{101, 184, 101}
\definecolor{darkgreen}{RGB}{0,128,0}
\definecolor{lightteal}{RGB}{121,199,210}
\definecolor{lightyellow}{RGB}{255,241,244}
\definecolor{darkyellow}{RGB}{255,166,50}
\definecolor{darkblue}{RGB}{0,63,92}
\definecolor{lightpurple}{RGB}{153,102,255}
\definecolor{red}{RGB}{178,34,34}
\definecolor{gray}{RGB}{166,166,166}
\definecolor{lightgreenhighlight}{RGB}{210, 235, 216}
\definecolor{lightredhighlight}{RGB}{252,172,174}

\definecolor{lightblue}{HTML}{DCF4FF}
\definecolor{darkblue}{HTML}{003F5C}
\definecolor{pink}{HTML}{C00072}
\definecolor{lightpink}{HTML}{FDD8ED}
\definecolor{darkpurple}{HTML}{631396}
\definecolor{darkyellow}{HTML}{AC7000}

\newcommand{\exactDot}{~\textcolor{pink}{\large$\bullet$}}
\newcommand{\relatedDot}{~\textcolor{darkblue}{\large$\bullet$}}

\newcommand{\topQuartile}{~\textcolor{darkyellow}{\large$\blacktriangle$}}

\newcommand{\shortmetaphor}[2]{\textit{{#1} is {#2}}}

\newcommand{\plainquote}[1]{{``#1''}}
\newcommand{\emphquote}[1]{\emph{``#1''}}

\newcommand{\etal}{{et~al.}~}

\begin{document}
\newcommand{\figMethods}{
    \begin{figure*}[h]
        \includegraphics[width=\textwidth]{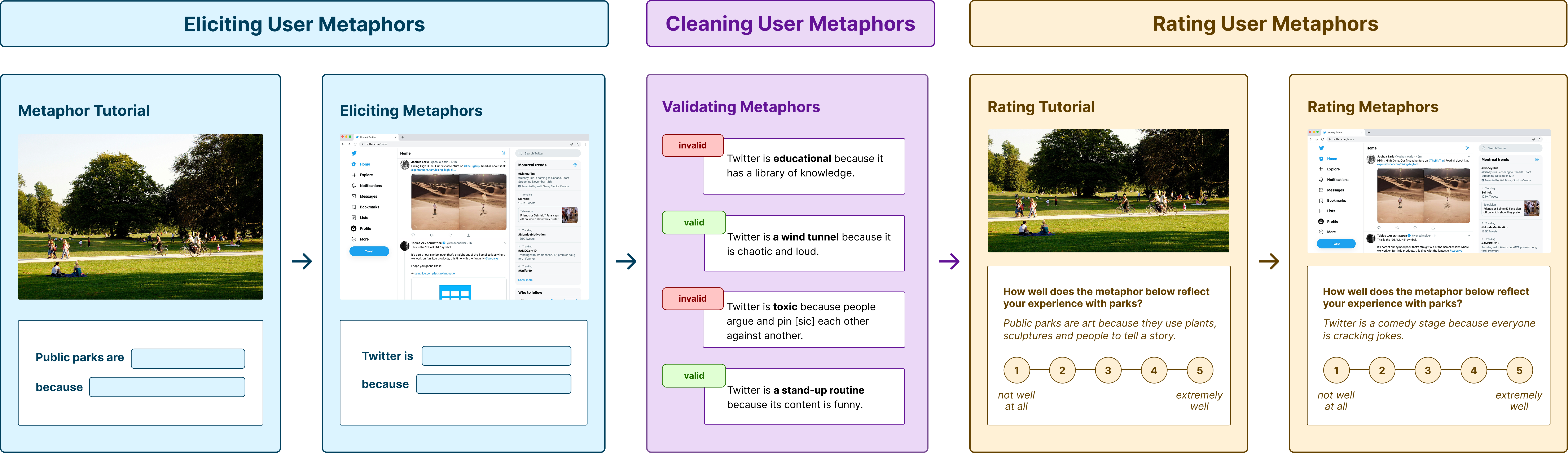}
        \caption{The user metaphor elicitation process. We extend prior sentence completion methods to elicit users' metaphors for their experiences with platforms. After removing invalid metaphors, we use the elicited metaphors to understand the experiences of a broader pool of participants.}
        \Description{A flow chart of the survey process for eliciting user metaphors.}
        \label{fig:methods}
    \end{figure*}
}

\newcommand{\figDesignerMetaphors}{
    \begin{figure}[t]
        \includegraphics[width=\columnwidth]{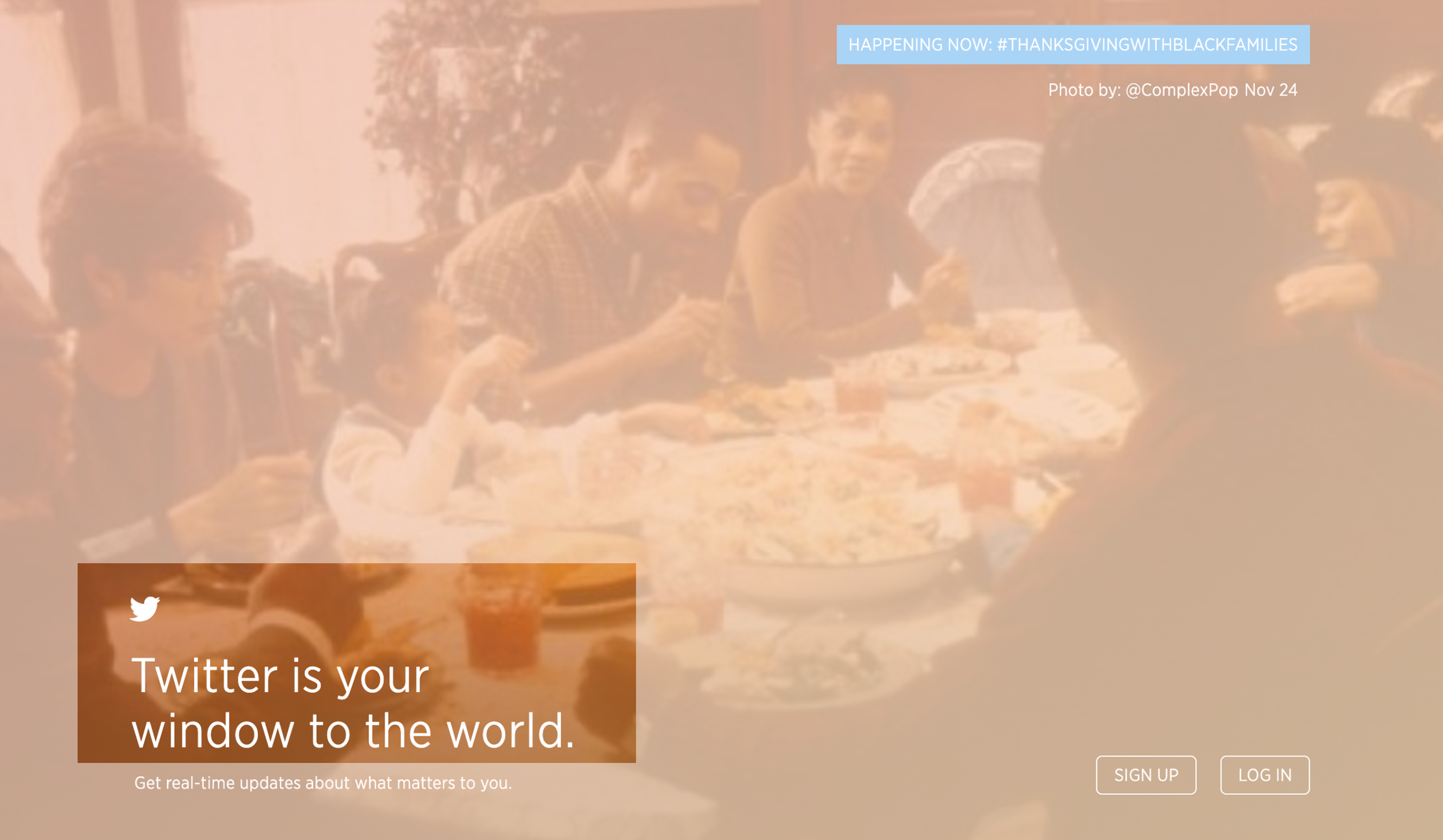}
        \caption{Twitter's about page in the first quarter of 2016. To identify design metaphors, we utilized the WayBack Machine to view the home and about pages of platforms from their launch date until 2024. We collected screenshots of both pages on the first day of the first and third quarter, when available. Next, we looked for \shortmetaphor{[PLATFORM]}{[METAPHOR]} statements to identify design metaphors.}
        \Description{Example of a surfaced design metaphor from Twitter's about page in 2016 suggesting that \textit{Twitter is your window to the world}.}
        \label{fig:designer-metaphors}
    \end{figure}
}

\newcommand{\figMatches}{
    \begin{figure*}[]
        \centering
        \includegraphics[scale=0.95]{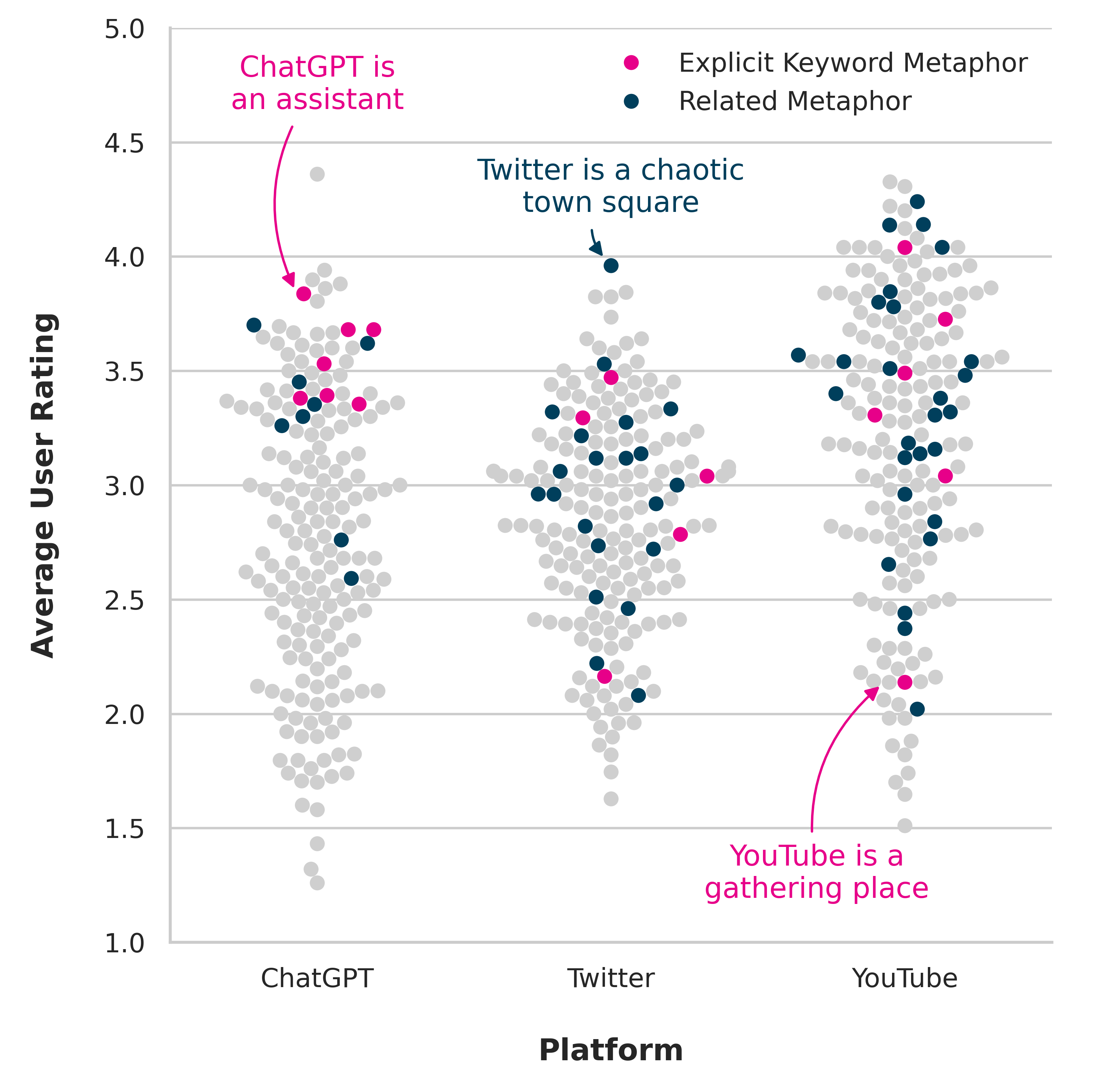}
        \caption{The average ratings for how reflective users found each user metaphor. We surveyed a broader, additional pool of participants to understand how reflective user metaphors' were of other users' experiences. Participants rated the reflectivity of metaphors on a 5-point Likert scale, with one being \textit{not well} and five being \textit{extremely well}. We average these ratings (approximately 50 ratings per metaphor) and indicate user metaphors that match or are related to design metaphors. We find that metaphors that contain explicit or related keywords from design metaphors are a mixed bag when it comes to how well they reflect users' experiences.}
        \Description{A comparison of three categorical swarm charts - one for ChatGPT, Twitter, and YouTube, showing the average rating of user metaphors for each platform based on participants' responses to "how well does this metaphor reflect your experience with the platform"? Participants responded using a 5-point Likert scale, in which 1 was not well at all and 5 was extremely well. We find that even when users' metaphors reflect or are related to design metaphors, they may not always reflect the wider experiences of users.}
        \label{fig:results-matches}
    \end{figure*}
}

\newcommand{\figcodes}{
    \begin{figure*}[]
        \centering
        \includegraphics[width=\textwidth]{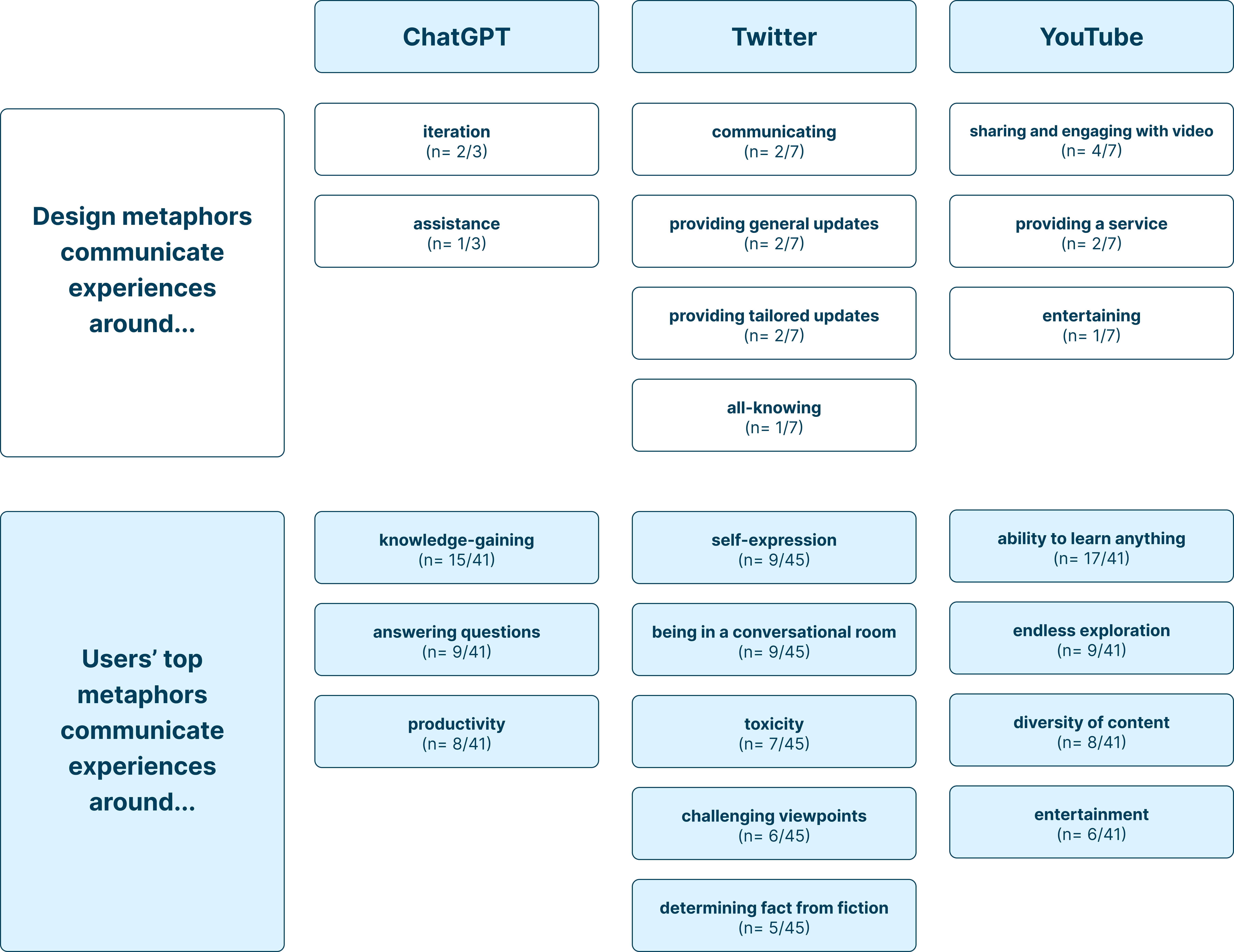}
        \caption{The first author and an independent annotator worked to group similar experiences described across design metaphors and the metaphors with average ratings in the top quartile. We include codes for the surfaced design metaphors and the most common codes for the top quartile of user metaphors for each platform (codes that apply to 5 or more user metaphors). The full set of codes for the top quartile of user metaphors are included in the supplementary material.}
        \Description{A comparison of user experiences---surfaced via qualitative coding---that are encapsulated by design metaphors and user metaphors.}
        \label{fig:results-codes}
    \end{figure*}
}

\title{Comparing Design Metaphors and User-Driven Metaphors for Interaction Design}

\author{Beleicia Bullock}
\email{beleicia@cs.stanford.edu}
\orcid{1234-5678-9012}
\affiliation{%
  \institution{Stanford University}
  \city{Stanford}
  \state{California}
  \country{USA}
}

\author{James A. Landay}
\email{landay@cs.stanford.edu}
\orcid{1234-5678-9012}
\affiliation{%
   \institution{Stanford University}
  \city{Stanford}
  \state{California}
  \country{USA}
}

\author{Michael S. Bernstein}
\email{msb@cs.stanford.edu}
\orcid{1234-5678-9012}
\affiliation{%
   \institution{Stanford University}
  \city{Stanford}
  \state{California}
  \country{USA}
}
\renewcommand{\shortauthors}{Bullock et al.}

\begin{abstract}
Metaphors enable designers to communicate their ideal user experience for platforms. Yet, we often do not know if these design metaphors match users’ actual experiences. In this work, we compare design and user metaphors across three different platforms: ChatGPT, Twitter, and YouTube. We build on prior methods to elicit 554 user metaphors, as well as ratings on how well each metaphor describes users’ experiences. We then identify 21 design metaphors by analyzing each platform’s historical web presence since their launch date. We find that design metaphors often do not match the metaphors that users use to describe their experiences. Even when design and user metaphors do match, the metaphors do not always resonate universally. Through these findings, we highlight how comparing design and user metaphors can help to evaluate and refine metaphors for user experience.

\end{abstract}




\begin{teaserfigure}
    \includegraphics[width=\textwidth]{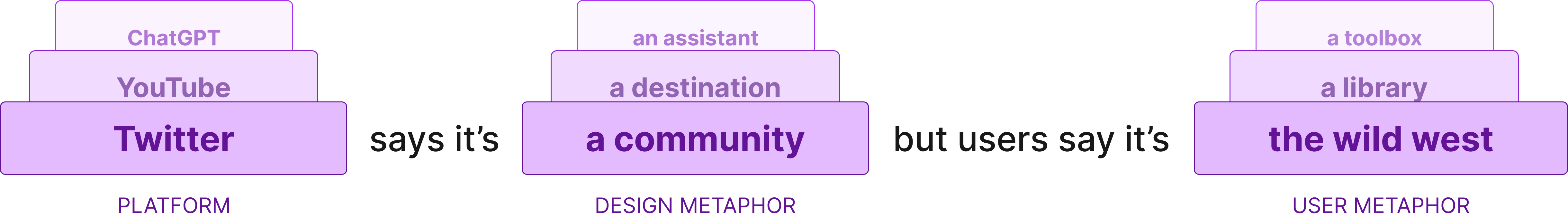}
    \caption{In this paper, we show how comparing the design metaphors that platforms' project against the metaphors that users experience can provide helpful insight for iterating design.}
    \Description{A graphic that compares design metaphors for each platform to user metaphors for each platform.}
\end{teaserfigure}

\maketitle

\section{Introduction}
As we look to develop positive and beneficial experiences with interactive systems, we often struggle to evaluate the effectiveness of the design metaphors that shape these experiences. Metaphors can be found in nearly all interactive systems~\cite{hsu_approach_2007} and they enable designers to communicate the capabilities and intended experience of such systems~\cite{sease_metaphors_2008, cila_generating_2020}. Given the impact of such metaphors, human-computer interaction and the broader design research community have developed a rich body of work that investigates what it takes to develop \plainquote{good} metaphors for interactive systems. However, as Sease~\etal notes, users are often \plainquote{left out of the communication process} when it comes to the generation and application of design metaphors~\cite{sease_metaphors_2008}. This creates several questions---namely \textit{do design metaphors actually translate to users and their experiences?} Are these metaphors reflective of all users' experiences, or only for specific groups? If design metaphors are not reflective of user metaphors, then which ones are, and what are the differences?

Given concerns about designers imposing metaphors onto users, prior work has tried to address this issue by exploring ways to incorporate user perspectives into conversations about what makes a good design metaphor. Early work in this area published guidelines that encouraged designers to define metaphor effectiveness based on users' actual behaviors~\cite{carroll_interface_1988, lovgren_how_1994}, to consider the cultural nuances of designing with specific metaphors~\cite{evers_cross-cultural_1998} and to evaluate the potential mismatches between the metaphor and the system~\cite{anderson_minimising_1994, carroll_interface_1988}. Later work shifted from providing practical guidance on incorporating users' perspectives to conducting research and developing tools that did. For example, Cila~\etal presents a more user-driven definition of metaphor quality by seeing how users rate the novelty, beauty, understandability, obviousness, and identifiability of product-metaphor pairs~\cite{cila_determinants_2014}. In addition, further work has looked to understand how metaphors' attributes impact non-designers' perspectives on aesthetics~\cite{cila_determinants_2014}. Notably, they found that participants were able to better ascertain the relationship behind product metaphors when the metaphors highlighted pragmatic or usability attributes and were created by expert designers. Such work created early steps towards defining the \plainquote{goodness} of metaphors through users' perspectives. Further work has also explored accessible tools for involving users in metaphor generation. Logler~\etal presents a step-by-step process for developing generative metaphors with input from stakeholders~\cite{logler_metaphor_2018}. Lockton~\etal outlines several activities that can be used to incorporate metaphorical insights into design research, including an activity that asks users to draw how they view concepts in order to generate new ideas for a domain~\cite{lockton_metaphors_2019}. Combined this research has worked towards addressing long-term calls for more user perspectives within the creation and development of design metaphors. However, we still lack methods to understand if design metaphors translate to users, how these translations differ across different communities, and how we can further center users in our development of design metaphors.

To this end, we already have tools that can be adapted to understand users' experiences with systems - namely, metaphor elicitation methods. While metaphor generation has often been seen as actively steered by designers, metaphor elicitation methods highlight how everyday people utilize metaphors to make sense of the world around them~\cite{lakoff_conceptual_1980}. Some of the earliest work in this field looked to surface customer perspectives through an interview-based method called the Zaltman metaphor-elicitation technique (ZMET)~\cite{christensen_mapping_2002, zaltman_marketing_2008}. Within HCI, we have seen the use of both interview-based and sentence-completion based metaphor elicitation methods. French \etal identified four dominant folk theories around feed-ranking algorithms for Facebook and Twitter\footnote{While the platform has been re-branded to X, we refer to the platform as Twitter in this paper and within our surveys.} by asking users to generate metaphors for how they believed such algorithms worked. More recent work by Cheng et al. applied a similar strategy using fill-in-the-blank style questions to understand how American users believed a broad class of AI systems worked~\cite{cheng_tools_2025}. In addition to understanding users’ perspectives on different types of systems, metaphor elicitation methods can also surface metaphors to understand user behavior. Seering~\etal constructed 22 moderation archetypes based on the metaphors moderators used to describe their philosophies, community roles, and users~\cite{seering_metaphors_2022}. Desai~\etal drew metaphors from both user interviews and prior literature to investigate how user metaphors indicate different communication modes with voice-based interfaces~\cite{desai_metaphors_2023}. Combined, this prior work shows that users are able to metaphorize their experiences with systems and do so in ways that generate deeper system insights~\cite{lockton_metaphors_2019}.

In order to evaluate the effectiveness of our metaphors, we propose leveraging these existing methods to compare user and design metaphors for specific systems. By comparing user metaphors to design metaphors, we not only get to understand users' perspectives and behaviors with systems, but also how their actual experience compares to the \textit{intended} experience shaped by designers and design metaphors. If a platform that is intended to feel like \plainquote{expressive writing partners}~\cite{openai_introducing_2025} is described as a \plainquote{magic 8-ball},  we learn that the \plainquote{writing partner} metaphor is not translating to users and receive some insight on how it is missing the mark (i.e., being formulaic or even toy-like). Moreover, if we get additional feedback from other users, we can understand how pervasive this \plainquote{magic 8-ball} experience is and if the experience varies across different communities. Such comparisons can enable feedback on the effectiveness of design metaphors and promote a more iterative and user-centered process for developing metaphors.

In this paper, we compare design and user metaphors across three platforms - ChatGPT, Twitter, and YouTube - to understand if and how design metaphors translate to users. We draw on prior fill-in-the-blank style methods to elicit \textit{user metaphors} for their experiences with a system~\cite{french_whats_2017, cheng_tools_2025}. We then collect a broader set of user ratings for each metaphor to understand how the metaphors reflect broader and community-specific experiences across each platform. We compare these user metaphors to \textit{design metaphors}, which we collect by analyzing the metaphors that platforms have published on their websites since their launch dates. Through this comparison, we identify design metaphors that translate to users, provide a comparison of the experiences that shape user and design metaphors, and examine of how these metaphors differ across different user demographics. With this work we contribute a new way to apply user-generated metaphors, a light-weight process to help designers collect feedback on their metaphors, and an understanding of how user metaphors can provide insight for shaping positive user experiences.
\section{Related Work}
We begin by defining what a metaphor is and how metaphors provide insight on people's experiences with the world around them. We then show how these qualities have been leveraged within human-computer interaction and broader design practice to support better design. Specifically, we highlight both the pragmatic uses of metaphors, which were applied to early graphical user interfaces, and the experiential uses of metaphor, which are explored in recent work. We show that while this literature has developed our understanding of what qualifies as a ``good'' metaphor, we lack an understanding of how such metaphors translate to users and their experiences in practice. We propose extending prior metaphor elicitation methods to fill this gap.

\subsection{Metaphors, Cognition, and Experience}
More than just a ``linguistic ornament''~\cite{moser_metaphor_2000}, metaphors are a ``critical cognitive device''~\cite{arora_typology_2012} that people use ``to think with, to explain themselves to others, [and] organize'' the world around them~\cite{gibbs_jr_metaphor_2008}. At a base level, a metaphor explains one concept in terms of another. As such, all metaphors are somewhat incorrect; otherwise they would be exact definitions~\cite{lockton_new_2019}. Despite these imperfections, we use metaphors everyday to make sense of our experiences and make decisions about life. Take the metaphor ``argument is war''. Here, the tenor - or target experience of an argument - is described by the vehicle - or source domain - of war~\cite{sease_metaphors_2008, logler_metaphor_2018}. Through this canonical example, Lakoff and Johnson show how our metaphorical construction of war influences our actions (e.g., picking sides, taking stances), values (e.g., winning, having good strategies), and experiences during arguments (e.g., with the same hostility and positioning as war)~\cite{lakoff_metaphors_2003,gibbs_jr_metaphor_2008}. By drawing on concepts that we are intimately familiar with, metaphors help us to understand new and/or more abstract experiences.

Given their grounding in the familiar, metaphors are shaped by our various contexts. As Moser indicates, metaphors are molded by cultural and societal norms, historical shifts, and idiosyncrasies~\cite{moser_metaphor_2000}. As such, a person who is raised in a culture where arguments are related to dance or negotiations~\cite{lakoff_conceptual_1980} will likely conceptualize arguments differently than a person raised in a culture where arguments are associated with war. Each metaphor helps the person to make sense of argumentation, but also creates challenges during interactions with individuals who have different conceptualizations. These resulting cognitive, contextual, and ubiquitous impacts have made metaphors a useful point of intervention for several fields, including education, psychology, and even interaction design. For instance, recent work has explored how metaphors shape public debates on emerging technology, such as on artificial intelligence (AI). Here, metaphors can support particular narratives~\cite{gilardi_we_2024} and affect how AI information is filtered through different political lenses~\cite{yang_ai_2025}.

\subsection{Design Metaphors in HCI and Design}
While metaphors where first applied in HCI to natural language systems, they gained prominence through the advent of graphical user interfaces (GUIs)~\cite{sease_metaphors_2008, hsu_approach_2007}. At the time, GUIs were a novel form of interaction and designers faced several constraints with deploying GUI-based systems. For one, most users would be novices. Moreover, GUIs were deployed in contexts that required reliability and effectiveness - namely workplaces~\cite{carroll_interface_1988}. To navigate these challenges, early interfaces drew inspiration from familiar desk tops - birthing the canonical metaphor and a broader research area around design metaphors~\cite{axtell_underdeveloped_2023, sease_metaphors_2008}.

\subsubsection{Employing Metaphors for Pragmatic Uses}
In particular, early work around metaphors in HCI focused on pragmatic uses of metaphor - that is, utilizing metaphor to describe and structure the systems' functionalities using concepts that users were familiar with~\cite{neale_role_1997}. These design metaphors helped to ``facilitate learning''~\cite{neale_role_1997} for novice users, manage the complexity of interfaces~\cite{carroll_interface_1988}, provide a ``useful model of the system''~\cite{erickson_working_1995} and ``increase the initial familiarity of actions''~\cite{carroll_interface_1988}. Subsequently, the field quickly adopted the use of metaphors in design~\cite{sease_metaphors_2008}, producing work around metaphor properties~\cite{carroll_metaphor_1985, anderson_minimising_1994}, theories~\cite{carroll_metaphor_1985}, frameworks~\cite{alty_framework_2000}, and a few guidelines~\cite{carroll_metaphor_1982, erickson_working_1995} to support designers in selecting interface metaphors. This structure naturally reflects the surface-level way that we often employ metaphors---``mapping features of an existing or familiar situation onto a new or unknown one'' to understand the unfamiliar better~\cite{lockton_new_2019}.
\subsubsection{Employing Metaphors for Experiential Uses}
However, as the notion of ``good'' design shifted from enabling task completion to producing broader, more positive experiential outcomes~\cite{law_understanding_2009, hassenzahl_needs_2010, hassenzahl_experience_2010, rivero_systematic_2017}, metaphors also were adapted to improve users' experiences. As Cila~\etal notes, metaphors enable designers to communicate their intended experience to users~\cite{cila_digging_2014}. These metaphors could target a number of experiential aspects including novelty, utility, feelings, and even statements~\cite{hekkert_handle_2015, cila_generating_2020}. As such, prior work explored what led to good metaphors for user experience. For example, prior work has studied how users' opinions on aesthetics change across the subtlety and abstractedness of metaphors~\cite{cila_product_2012}. Another study explored how the type of intentions communicated by a metaphor (e.g., highlighting how to use a product versus embedding a specific experience) and the expert level of designers impact the depth of resulting metaphors~\cite{cila_digging_2014}. While understanding metaphorical attributes and their impact is an important factor for choosing design metaphors, much of this work developed notions of a good metaphor without understanding if the intended experience was reflected in users' experiences. Specifically, does the experience of a ``town square'' translate to users of a social platform? If so, does it translate to all users or just some? Does it translate in totality (i.e., design, actions, behavior, values, etc.) or just in particular parts?

\subsection{User Metaphors in HCI}
While the work above does not address translation to users, we can extend recent work around metaphor elicitation to fill this gap. Namely, recent work has explored new, evaluative possibilities powered by user-generated metaphors. So far, these methods have shown that metaphors can help us to understand users' mental models and behaviors with systems. We look to build upon this research by expanding to a third domain - experience.

\subsubsection{Identifying Mental Models via User Metaphors}
One of the earliest works around user-generated metaphors showed how such metaphors provide insights around users' mental models of algorithmic feeds~\cite{french_whats_2017}. In particular, French~\etal developed an interactive wikisurvey with a set of seed metaphors to describe how Facebook and Twitter's feed ranking algorithms worked~\cite{french_whats_2017}. Participants could indicate how they thought the feeds worked by selecting one of the seeded metaphors or by entering their own. Through the resulting seeded and user-generated metaphors, French~\etal identified four dominant folk theories about algorithmic feeds based on users' metaphors. We still see the use of similar methods today---recent work by Cheng~\etal adopt similar methods to understand the metaphors Americans' use to describe how AI tools work and how these metaphors reflect different patterns of trust and warmth over time~\cite{cheng_tools_2025}. 

\subsubsection{Identifying Behaviors via User Metaphors}
In addition to mental models, users' metaphors have also been used to classify users' behaviors and interactions with systems via qualitative interviews and literature reviews. For example, Seering~\etal~\cite{seering_metaphors_2022} interviewed over 70 moderators of online Twitch, Reddit, and Facebook Group communities to understand the metaphors for their moderation practices. These interviews resulted in 5 moderation modes and 22 archetypes surfaced from the metaphors that moderators used to describe their actions, community roles, and interactions with users. Exploring behavior across a different set of technologies, Desai~\etal explored the use of different user metaphors for voice-based user interfaces across prior literature, product design and user interviews~\cite{desai_is_2022,desai_metaphors_2023}. From this analysis, they identified four metaphors for how users instruct voice-based user interfaces and how these metaphors varied across the same user depending on the context. Together, these works articulate the potential for user-generated metaphors to inform design. We build on this work by exploring another domain - understanding users' experiences. In this project, we extend this prior work by asking how effectively design metaphors are based on their translation to users' metaphors. To do this, we draw on theory, definitions, and methods from prior work---for example, developing a variant of prior metaphor elicitation methods~\cite{french_whats_2017, cheng_tools_2025}---and apply those ideas to compare design metaphors and users' metaphors for the same system. 
\section{Methodology}
\label{methods}
For the purposes of this paper, we define ``design metaphor'' as the metaphor projected outwards by the artifact, and ``user metaphor'' as the metaphor understood by users. In order to understand how design metaphors translate to users and across different user communities, we surface and compare design and user metaphors across three existing systems - ChatGPT, Twitter, and YouTube. Such systems cover a wide design space (e.g., the networked presentation of Twitter versus the commons model of YouTube or the individualized, agentic presentation of ChatGPT~\cite{zhang_form-_2024}), purposes, histories and user communities, allowing us to explore how comparing design and user metaphors can surface varied insights. We use the following sections to indicate how we define and surface each type of metaphor, as well as our methods for comparing the two.
\subsection{Surfacing Design Metaphors}
\label{methods:designer-metaphors}
\figDesignerMetaphors
While prior work has highlighted the importance of understanding design metaphors, identifying such metaphors is a non-trivial challenge~\cite{desai_metaphors_2023}. For example, Desai~\etal notes the lack of manufacturer perspectives for voice-based user interfaces within the ACM Digital Library~\cite{desai_metaphors_2023}. Dove~\etal takes a more ethnographic approach to surfacing design metaphors by studying the emergence and development of the metaphor ``planting a seed'' across the design of an interactive art piece~\cite{dove_life_2018}. While this provides valuable information on how a design metaphor may transform and inform design over the course of a project, we lack methods that help us to surface such metaphors after deployment - creating a bottleneck for studying the evolution and impact of design metaphors beyond generation.

Given these challenges, we draw inspiration from discourse analysis methods used in communication critiques of existing platforms. In particular, these methods examine discourse around systems to surface the underlying politics, logics, motivations, and values that shape systems. Prior work uses such methods to analyze the language exhibited in the social media posts of platform executives~\cite{hoffmann_making_2018}, company blog posts and announcements~\cite{christin_internal_2024}, and instances of the \emphquote{platform} metaphor across companies' interactions with users, media, and policy makers~\cite{gillespie_politics_2010}. In particular, we see parallels between Hoffmann~\etal's self-definitions - or the language used by stakeholders to define platforms (e.g., [PLATFORM] is...) - and the ways in which metaphors explain concepts in terms of another~\cite{hoffmann_making_2018}.

As such, we define design metaphors as metaphors presented by platforms in the form of ``[PLATFORM] is [METAPHOR]''. Specifically, we look for such metaphorical self-definitions on the home and about pages of platforms' websites. Both pages provide companies with space to present their platform to the user and provide descriptions of the platform. While platforms may also contain text that allude to more implicit metaphors~\cite{desai_metaphors_2023}, such metaphors could require additional interpretation that may veer from the platform's intended meaning.

Our method reflects the reality that the metaphors projected outwards by platforms may be crafted by a wide variety of people at an organization. So why call these ``design metaphors''? While we acknowledge that discourse around platforms can come from a variety of stakeholders, we argue that these stakeholders all impact design outputs even if they do not carry the traditional label of ``designer''. For one, design itself is a multi-stakeholder, ``social process of collaborative inquiry''~\cite{dove_life_2018}. All stakeholders - whether they are designers, engineers, managers, marketers, or policy makers - must engage with one another in order to produce design outputs~\cite{schon_designing_1988}. Furthermore, prior work shows that corporate design work is often expansive, collaborative and multi-disciplinary~\cite{kang_challenges_2024, kou_towards_2018}. Designers' perspectives are both shaped by the perspectives of practitioners in ``overlapping occupational domains (e.g., developers, marketers, visual designers)''~\cite{kou_towards_2018} and conveyed to relevant stakeholders through discourse~\cite{wong_tactics_2021, kang_challenges_2024}. Reflecting this reality, a recent analysis of over 3,000 design-related job posts indicates that deliverables for UI and UX positions range from traditional artifacts such as webpages and prototypes to non-interaction artifacts, including social media content, campaigns, and content strategy~\cite{xu_ux_2025}. As such, even material that appears more marketing-related may reflect input from traditional designers and generate insights on how platforms were designed. In sum, this collaborative and expansive view of corporate design suggests that discourse surfaced from the home and about pages can provide a good approximation of the metaphors that shaped a platform's design.

\subsubsection{Procedure}
To surface design metaphors, we analyze text across screenshots of each platform's home and about pages from their launch data up until 2024. We assembled our corpus of screenshots using the \href{https://archive.org/}{Wayback Machine} - an archival project from the Internet Archive that contains over 900 million snapshots of web pages. Both Twitter and YouTube launched in the early 2000s, while ChatGPT was released for public use in 2022. Since designs likely do not change multiple times a year, we take screenshots on January 1st (Q1) and July 1st (Q3) of each year up to 2024 (when our user metaphors were collected). If the archived version of the page was not available that day, the archived version of the page was not in English, or the only archived version of the page was a redirect (due to changes in URLs and/or platforms' sitemaps), we collect the screenshot from the first available day within that specific quarter. For pages that had interactive elements like slide shows, we took screenshots of each individual state and combined them together to form one screenshot.

Once all of the screenshots were collected, we reviewed each screenshot to find any candidate metaphors publicized on the site. Drawing on Hoffman's format for platforms' self-definitions~\cite{hoffmann_making_2018}, a candidate metaphor is defined as a statement in the form [PLATFORM] is [BLANK] or a statement that could be placed in this form. This included reviewing taglines, press quotes placed on the homepage, and any text explaining the platform. However, we did not include text from user-generated content, such as tweets or YouTube video titles. We show an example of what a candidate metaphor looks like in Figure~\ref{fig:designer-metaphors}, which shows Twitter's home page on January 1st, 2016. We consider the headline - ``\textit{Twitter is your window to the world}'' - to be a candidate metaphor. However, other taglines from Twitter such as ``\textit{Let your imagination run wild}'' and \emphquote{Empower your entire workforce} would not be considered candidate design metaphors because they cannot be converted into [PLATFORM] IS [BLANK] statements and do not constitute a clear definition of the platform. The tagline \emphquote{Empower your entire workforce}, for instance, may suggest implicit metaphors of \textit{energy}, \textit{power}, or even \textit{support}. However, without an explicit metaphor, it is unclear if these metaphors actually reflect how Twitter's stakeholders would conceptualize such a statement.

Through this process, we identified 33 candidate metaphors. We then reviewed each metaphor to see if the statement was a valid metaphor. Validity was determined by explicitly relating the platform to a place, object, person, or concept. While this is not the only way to define metaphors, it does ensure that there is an explicit comparison being made and reduces the potential issues of interpreting an implicit metaphor. For example, we do not count the statement ``\textit{YouTube is building a community that is highly motivated to watch and share videos}'' as a metaphor because it is not compared to a place, object, person or concept. Moreover, it is unclear if YouTube is actually being framed as a community or community builder. This refinement resulted in a total of 21 design metaphors - 3 metaphors for ChatGPT, 10 for Twitter, and 7 for YouTube.

\subsection{Surfacing User Metaphors}
\figMethods
In contrast to methods for design metaphors, several methods have been proposed for surfacing user metaphors. Specifically, we look to elicit metaphors that reflect the overall experience of users, rather than surfacing mental models of how broad classes of systems work~\cite{french_whats_2017, cheng_tools_2025} or synthesizing specific user behaviors~\cite{desai_metaphors_2023, lockton_metaphors_2019}. As such, we extend prior methods to develop a two-part survey that elicits (1) explicit user metaphors to describe users' experiences via sentence completion techniques~\cite{lallemand_optimizing_2022, schmidt_sentence_2024, walsh_exploring_2010, french_whats_2017, cheng_tools_2025} and (2) user consensus across these metaphors via ratings~\cite{wobbrock_user-defined_2009, morris_reducing_2014}.

\subsubsection{Elicitation Participants}
\label{methods:elicitaion-participants}
To elicit user metaphors, we identify, recruit, and compensate participants using the survey platform, Prolific. We recruit 100 participants per platform for the elicitation portion of our study, resulting in a total of 300 recruited participants. Participants had to be located in the United States, proficient in English, and users of the respective platform as indicated by their Prolific profiles. Furthermore, participants could not have participated in prior pilots of the study or in the surveys for the other two platforms. Participants were compensated at a pro-rated rate of \$20 an hour and compensation was based on answering at least one attention check correctly, submitting the survey and entering the correct completion code. If the participant failed both of the attention checks or did not complete the survey, they received a code that returned their survey but did not jeopardize their standing on Prolific. From August to September 2024, 108 participants attempted our YouTube elicitation survey (99 completed, 6 returned, 2 timed out, 1 incomplete), 110 attempted our ChatGPT elicitation survey (100 completed, 8 returned, 2 timed out), and 102 attempted our Twitter elicitation survey (100 completed, 1 returned, 1 timed out).

\subsubsection{Elicitation Procedure}
\label{methods:elicitation-survey}
Despite being part of our everyday lexicon, generating metaphors is not easy for designers, nor everyday users~\cite{lockton_metaphors_2019}. As such, we first presented participants with a training section before eliciting metaphors about their experiences with platforms. Specifically, we asked participants to consider their experiences in a public park at the start of our Qualtrics survey. We presented them with a stock image of a park and encouraged reflection by delaying the display of the ``Next'' button for five seconds. Once the button appeared, participants were shown four potential metaphors to describe their experiences with parks: \textit{lungs}, \textit{oases}, \textit{community centers}, and \textit{classrooms}. Each metaphor used the structure ``Public parks are [METAPHOR] because [EXPLANATION].'' The order of the metaphors was randomized for each user. Participants were able to select one or multiple metaphors to best capture their experiences. Participants were then asked to generate their own metaphor-reason pair for their experiences with parks using fill-in-the-blank text boxes (as seen in Figure~\ref{fig:methods}).

After this training section, we asked participants to generate metaphors for the platform featured in the survey. We followed a similar format to the training activity above: we presented participants with an image of the target platform, asked participants to reflect on their past experiences, and then generate a metaphor-reasoning pair for their experience using fill-in-the-blanks. However, instead of asking for one metaphor, we asked participants to provide two. This multiplicity has been suggested as a way to combat legacy bias - or the tendency to default to conventional representations and metaphors~\cite{morris_reducing_2014}. While conventional metaphors might be reflective of societal or cultural norms, they may not reflect users' true, individual thoughts. 

For similar reasons, we also provided examples of metaphors for a technical artifact  - a digital camera. Namely, we included metaphors drawn from places (e.g.: \textit{an art studio}), objects (e.g.: \textit{a mirror}), and concepts (e.g.: \textit{story time}) to show the breadth of real-life ideas that users could draw from. Providing examples is also a suggested practice for combating legacy bias and we aimed to do this in a way that did not have overlap with our platforms~\cite{morris_reducing_2014}. 

Finally, we conclude the survey with a demographic survey that asked users about their platform usage, race, gender, and age. While self-reported usage can be inaccurate, using more accurate time-tracking solutions were outside the scope of our study. As such, we utilized a self-reported usage scale that has been shown to have lower error rates than other methods~\cite{ernala_how_2020}. Specifically, we asked users to indicate their usage over the past week using the following time range: less than 10 minutes per day, 10-30 minutes per day, 31-60 minutes per day, 1-2 hours per day, 2-3 hours per day, more than 3 hours per day, and didn't use. Beyond usage, we looked to support users' autonomy by only requiring users to indicate their age and including optional entries for race and gender. When asking about race, we drew from Stanford University's IDEAL survey questions, which offered expanded race and ethnicity categories~\cite{stanford_university_survey_nodate}. We include a template of the survey in the supplementary materials.

Through this process, we elicited 598 candidate user metaphors. Much like our candidate design metaphors, not all candidate user metaphors resulted in explicit and/or valid metaphors. For example, \plainquote{ChatGPT is eerie because it holds onto all of your request [sic] that are able to be resurfaced later} may contain an implicit metaphor (such as ChatGPT being a bad feeling) but does not contain an explicit metaphor. Another common example of non-explicit metaphors were definitions such as \plainquote{ChatGPT is technology} or \plainquote{Twitter is a social platform}. Beyond these instances, we removed handful of candidate metaphors that 1) presented potentially harmful text~($n=2$) due to the use of slurs or problematic metaphors like ``slave''~\cite{desai_metaphors_2023}, 2) were unclear to the authors due to spelling~($n=2$), and 3) were potentially generated by an LLM based on their similarity to generated responses from ChatGPT when prompting for platform metaphors~($n=2$). Once non-valid metaphors were removed, we converted similes to metaphors (e.g., \plainquote{like an open world} became \plainquote{an open world}), added in missing articles, corrected small spelling mistakes, and applied capitalization where necessary. Our final set of user metaphors contained 554 metaphors with 184 metaphors for ChatGPT, 188 for Twitter, and 182 for YouTube. 

\subsubsection{Ratings Participants}
\label{methods:ratings}
In addition to collecting user metaphors, we also surveyed a broader set of users to understand how well each metaphor reflected other users' experiences with platforms and how these experiences differed across communities.

\label{methods:ratings-participant}
We again used Prolific to recruit additional participants to rate how well each metaphor generated in Section~\ref{methods:elicitation-survey} reflected their experiences with the respective platform. We recruited three separate participant pools - one for each platform - from a pool of Prolific users that were located in the United States, had indicated using the respective platform in their Prolific profile, and had not participated in the elicitation surveys or the rating surveys for other platforms in our study. Furthermore, to ensure our ratings reflected the experiences of a diverse selection of participants, we included the following quotas: 50\% of our participants should identify as female (sex was the available filter for this) and there should be an equal quota across the available racial categories (in this case a 20\% split across Asian, Black, Mixed, White, and Other). We note that participants had more inclusive options for indicating (or not indicating) their race and gender. Participants were compensated for their time at a pro-rated rate of \$20 per hour. Their compensation was contingent upon passing at least one attention check, completing the survey, and submitting the completion code. The participants who attempted the survey but did not complete it or failed both attention checks were not compensated, but were asked to return their survey. In January 2025, we recruited 307 participants to rate user metaphors for ChatGPT, 314 to rate metaphors for Twitter, and 305 to rate metaphors for YouTube.

\subsubsection{Ratings Procedure}
\label{methods:ratings-survey}
To collect our metaphor ratings, we utilized Qualtrics surveys - one for each platform - to randomly present a selection of 30 user metaphors to each participant. We began the survey with a similar training activity as the one in Section~\ref{methods:elicitation-survey}. Participants were presented with images of a park and provided with time to reflect upon their prior experiences. Next, participants were presented with three different metaphors to describe their experiences with parks - a community center, medicine, and art. Instead of generating metaphors, we asked participants to use a 5-point Likert scale to rate how well each metaphor reflected their experience with parks. This scale contained the following ratings: \textit{not well at all}, \textit{slightly well}, \textit{moderately well}, \textit{very well}, and \textit{extremely well}.

Once participants completed this training activity, they were presented with 30 metaphors randomly and evenly selected from the user metaphors collected in Section~\ref{methods:elicitation-survey}. Participants ended the survey by completing a set of demographics questions around prior platform usage, race, gender, and age. 

\subsection{Metaphor Analysis and Comparison}
\label{methods:analysis}
\subsubsection{Metaphor Analysis}
\label{methods:metaphor-analysis}
As highlighted in prior work~\cite{desai_metaphors_2023,cila_generating_2020}, there a numerous ways to analyze and categorize metaphors. To analyze both design and user metaphors, the first author worked with an independent annotator with prior qualitative coding experience to group similar metaphors. Each annotator grouped and labeled metaphors separately, met to discuss disagreements, and compiled a final list of groups for both designer and user metaphors. Notably, we grouped metaphors by similar metaphors and explanations, rather than grouping on metaphors alone. For example, two user metaphors may relate their platform experience to that of a library. However, one user may utilize the library metaphor to communicate the amount of content on the platform (e.g., \plainquote{YouTube is a library because it offers content on every topic}), while another user utilizes the same library metaphor to articulate their ability to learn on the platform (e.g. \plainquote{YouTube is a library because you can learn almost anything there}). Similarly, design metaphors may communicate similar intended experiences using two different metaphors. The design metaphors \plainquote{YouTube is a personal video sharing service} and \plainquote{YouTube is a worldwide video-sharing community} draw on a similar experience (in this case being a medium for sharing) while conceptualizing this experience as two different metaphors. By looking at the entire metaphorical statement, we are able to better understand the similar design and user metaphors and differentiate the underlying experiences. 

\renewcommand{\arraystretch}{1.2}
\begin{table}[]
    \caption{The design metaphors for each platform. We surfaced 21 design metaphors by analyzing the screenshots of the home and about pages for each platform's website from its launch date to 2024. We indicate design metaphors with explicit keywords that are found in at least one user metaphor with a pink dot~\exactDot. Design metaphors with at least one related user metaphor are indicated with a dark blue dot~\relatedDot.}
    \label{tab:designer-metaphors}
    \small
    \begin{tabular}{p{0.95\columnwidth}}
    \textbf{ChatGPT is...} \\ \hline
    ...a \textbf{sibling} model to InstructGPT which is trained to follow an instruction in a prompt and provide a detailed response.\\
    ...the latest \textbf{step} towards the deployment of increasingly safe and useful AI systems.~\relatedDot\\
    ...a super\textbf{assistant} for every member of your team.~\exactDot~\relatedDot\\ \\

    \textbf{Twitter is...}\\ \hline
    ...a global \textbf{community} of friends and strangers answering one simple question: what are you doing?~\exactDot~\relatedDot\\
    ...the \textbf{telegraph} system of Web 2.0.\\
    ...is a \textbf{service} for friends, family, and co-workers to communicate and stay connected through the exchange of quick, frequent answers to one simple question: what are you doing?~\exactDot~\relatedDot\\
    ...a simple \textbf{tool} that helps connect businesses more meaningfully with the right audience at the right time.~\relatedDot\\
    ...a \textbf{source} of instant information because you can stay updated and keep others updated.~\exactDot~\relatedDot\\
    ...a real-time information \textbf{network} powered by people all around the world that lets you share and discover what’s happening now.~\relatedDot\\
    ...your \textbf{window} to the world because you get real time updates that matter to you.~\exactDot\\
    ...a real-time information \textbf{network} that connects you to the latest information about what you find interesting.~\relatedDot\\
    ...a real-time information \textbf{network} connects you to the latest stories, ideas, opinions and news about what you find interesting.~\relatedDot\\
    ...\textbf{ESP} (extrasensory perception).\\ \\

    \textbf{YouTube is...}\\ \hline
    ...the Internet’s premier video \textbf{service}.~\exactDot~\relatedDot\\
    ...a distribution \textbf{platform} for original content creators and advertisers large and small.~\relatedDot\\
    ...an entertainment \textbf{destination}.~\relatedDot\\
    ...an online video \textbf{community}, allowing millions of people to discover, watch and share originally-created videos.~\exactDot~\relatedDot\\
    ...a worldwide video-sharing \textbf{community}\\
    ...a personal video sharing \textbf{service}~\exactDot~\relatedDot\\
    ...a \textbf{place} for people to engage in new ways with video by sharing, commenting on, and viewing videos~\exactDot~\relatedDot\\
    \end{tabular}
\end{table}

\subsubsection{Metaphor Comparison}
\label{methods:metaphor-comparison}
To compare design and user meta\-phors, we utilize keywords to identify (1) user metaphors with explicit keywords from design metaphors and (2) user metaphors that are related to design metaphors. Given that design metaphors can vary in detail~(see Table~\ref{tab:designer-metaphors}), we look to simplify design metaphors to their core ideas via keywords. To do this, we select the simplest metaphor for each design metaphor as our explicit keyword. For example, \plainquote{Twitter is a global community of friends and strangers answering one simple question: what are you doing?} simply becomes \plainquote{Twitter is a community}, with \textit{community} as our explicit keyword. We then use the full metaphorical statement to identify user metaphors that may be related to design metaphors, but do not use the exact, same metaphor to describe their experiences. For example, a community may be a club for one user or a neighborhood for another. As such, the first author worked alongside an independent annotator with experience in qualitative methods to identify user metaphors that were similar to the full design metaphor, in addition to the explicit keyword from design metaphors. The pair independently reviewed all user metaphors for similarities with design metaphors and then met to discuss disagreements.

\subsubsection{Rating Analysis}
\label{methods:ratings-analysis}
Finally, we analyze users' metaphor ratings across three demographic areas: race, gender, and age. Typically, we would do this by running ANOVA and subsequent t-tests across the metaphors to identify pairwise differences between two demographic groups.

However, this approach runs the risk of a family-wise error, where so many statistical tests are being run that we might produce false positives even if each individual test is at $\alpha=0.05$. We are potentially running hundreds of these comparisons, since we want to test each metaphor separately across different racial, age, and gender groups. The appropriate statistical approach for this situation, characterized by large numbers of tests, is to control for false discovery rate (FDR). Specifically, we use the Benjamini-Hochberg correction~\cite{benjamini_adaptive_2000}, which is commonly used in fields such as genetics and biochemistry where large numbers of comparisons are tested. The Benjamini-Hochberg correction corrects a large number of tests to maintain an overall false discovery rate set by the researcher, which we set at $0.05$. We perform the Benjamini-Hochberg correction on each collection of metaphors for a single platform, for each demographic group. For instance, we test for differences in all YouTube metaphors by race. We only report significant differences if the omnibus ANOVA is significant, and if the pairwise t-test is significant after adjustment by the Benjamini-Hochberg correction.

\section{Results}
\textit{Do design metaphors translate to users' experiences?} To answer this question, we first present the design and user metaphors surfaced by the methods above. We then compare both sets of metaphors for each platform, indicating user metaphors that reflect explicit keywords from design metaphors or are related to design metaphors. We find that many of the user metaphors surfaced do not reflect design metaphors' explicit keywords and are not related to design metaphors. As such, we look to understand the experiences reflected across user metaphors by qualitatively grouping the top rated metaphors for each platform based on communicated experience. Finally, we explore metaphors that are rated differently across various user communities, including user metaphors that are rated in the top quartile of metaphors for their respective platforms.
\subsection{Design Metaphors}
\label{results:designer-metaphors}

We identified 21 design metaphors across all three platforms using the methodology outlined in Section~\ref{methods:designer-metaphors}. As seen in Table~\ref{tab:designer-metaphors}, we identified 10 metaphors for Twitter, 7 metaphors for YouTube, and 3 metaphors for ChatGPT. We see these metaphors converge around key platform characteristics, painting a picture of the intended experience of each platform. For instance, Twitter has several information and updates-related metaphors (e.g., \plainquote{telegraph system}, \plainquote{real-time information network}, \plainquote{window to the world}) that highlighted intended experiences around timely updates on current events and personalized updates on topics that matter to users~(Figure~\ref{fig:results-codes} displays all of the intended experiences qualitatively surfaced from design metaphors). In contrast, the majority of the design metaphors for YouTube (including \plainquote{video-sharing community}, and \plainquote{entertainment destination}) underscore an intended experience characterized by sharing and engaging with video. ChatGPT had the fewest design metaphors of all three platforms. Two of these metaphors (\plainquote{sibling model to InstructGPT}, \plainquote{the latest step toward the deployment of \ldots AI systems}) reflect intended experiences shaped by iteration. Beyond this, the only other design metaphor we surfaced for ChatGPT related the platform's intended experience to that of a \emphquote{superassistant}.

\subsection{User Metaphors}
\label{results:user-metaphors}
Through our metaphor elicitation process from Section~\ref{methods:elicitation-survey}, we surfaced 554 valid metaphors across all three platforms. These metaphors ranged from the foreseeable (e.g., \plainquote{ChatGPT is intelligence}) to the unexpected (e.g.,\plainquote{ChatGPT is a Taco Bell meal because it's awesome right now but you're gonna pay for it later}). We include all elicited, user metaphors in the supplementary materials.

\begin{table}[!ht]
    \caption{Top 5 metaphors for platforms based on the average rating of users' responses to \textit{``How well does [metaphor] reflect your experience with [platform]?}''. Users rate the reflectiveness of the each metaphor using a 5-point Likert scale ranging from \textit{not well} to \textit{extremely well}. While there are no user metaphors with explicit keywords from design metaphors in the top five rated metaphors for any platform, some user metaphors that are related to design metaphors do rank in the top five. We indicate these metaphors using a dark, blue dot~\relatedDot.}
    \label{tab:user-metaphors}
    \smaller
    \begin{tabular}{p{0.6\columnwidth}ccc}
    \toprule
    \textbf{User Metaphor} & \textbf{Mean} & \textbf{Std. Error}\\
    \toprule
    ChatGPT is a \textbf{tool} because you can use it for work & 4.36 & 0.13\\
    ChatGPT is an \textbf{encyclopedia} because it can be used to research tons of information & 3.94 & 0.14\\
    ChatGPT is a \textbf{toolbox} because it can help you solve questions and issues if you know how to use it & 3.90 & 0.15 \\
    ChatGPT is a \textbf{library} because it contains and collects a vast amount of information in one place & 3.88 & 0.16\\
    ChatGPT is a \textbf{library} because it provides vast amounts of knowledge & 3.86 & 0.15\\ \midrule
    Twitter is a \textbf{chaotic town square} because everyone can communicate out their ideas but it is difficult to tell fact from fiction~\relatedDot & 3.96 & 0.16\\
    Twitter is \textbf{the wild west} because people feel they can say anything online & 3.84 & 0.14\\
    Twitter is a \textbf{madhouse} because there are so many debates and arguments, especially about politics & 3.82 & 0.17\\
    Twitter is a \textbf{shield} because people will say things they won’t face to face & 3.82 & 0.17\\
    Twitter is a \textbf{rabbit hole} because there are many different threads of information to go down & 3.73 & 0.15\\ \midrule
    YouTube is a \textbf{library} because it houses millions of videos containing various types of content & 4.33 & 0.12 \\
    YouTube is a \textbf{never-ending abundance of learning opportunities} because so many ideas, lessons, and tutorials are available & 4.31 & 0.10\\
    YouTube is \textbf{entertainment} because it allows me to escape to a different place and experience new things~\relatedDot & 4.24 & 0.11\\
    YouTube is a \textbf{library} because you can access information and experiences about almost anything & 4.22 & 0.13\\
    YouTube is a \textbf{library} because there is a seemingly endless supply of entertainment and information & 4.20 & 0.15\\ 
    & & \\
    & & \\
    \end{tabular}
\end{table}

In Table~\ref{tab:user-metaphors}, we present the five user metaphors with the highest average rating for each platform. These results reflect some of the themes we saw across design metaphors (Table~\ref{tab:designer-metaphors}). For example, we see the iteration theme reflected in ChatGPT's design metaphor---\plainquote{ChatGPT the latest step towards the deployment of increasingly safe and useful AI systems}---reflected in a user metaphor that reflects evolution---\plainquote{ChatGPT is the future}~($\mu=3.70, SEM=0.16$). In other instances, we see completely diverging metaphors. While many of Twitter's design metaphors centered around information, we see that the metaphor that was most reflective of users' experience, on average, was \plainquote{Twitter is a chaotic town square because everyone can communicate out their ideas but it is difficult to tell fact from fiction}~($\mu=3.96, SEM=0.16$). Furthermore, information-rich experiences are only reflected in one of the metaphors in the top ten metaphors for the platform - \plainquote{Twitter is a rabbit hole because there are many different threads of information to go down}. We compare design and user metaphors in the next section to contextualize these results.

\subsection{Comparing Design Metaphors and User Metaphors}
\label{results:match}
\figMatches
\textit{Do design metaphors actually translate to users' experience?} We show that few user metaphors contain the explicit keywords from design metaphors. Thus, we broaden our search to include user metaphors that are related to design metaphors, but may not use the exact same language. We explore both results below.

\subsubsection{User Metaphors with Explicit Keywords from Design Metaphors}
We find that majority of user metaphors do not contain the explicit keywords referenced in design metaphors, as highlighted in Figure~\ref{fig:results-matches}. Of the 16 unique explicit keywords surfaced from design metaphors (see Table~\ref{tab:designer-metaphors}), eight of the explicit keywords where referenced in user metaphors. However, half of these keywords had fewer than 3 instances were user metaphors referenced them. Moreover, referencing an explicit keyword did not guarantee an exact match between experience expressed by the design metaphor and the experience expressed by the user metaphor. For example, the design metaphor \plainquote{Twitter is a \textbf{window} to your world} contains the explicit keyword \plainquote{window}. We surfaced the user metaphor \plainquote{Twitter is a curtain over a \textbf{window} on a sunny day because its user base puts a damper on what otherwise could be a great resource} using the explicit keyword. However, there is notable difference in the portal-like experience conveyed by the design metaphor and the experience of disappointment exhibited by the surfaced user metaphor. Although majority of our user metaphors do not contain explicit keywords found in design metaphors, we do note that each platform has at least one surfaced user metaphor that references an explicit keyword.

Beyond looking for explicit keywords from design metaphors across our corpus of user metaphors, we also consider how highly users rated metaphors that contained these keywords. Even if there only a few user metaphors that mention explicit keywords, other users' may have found the user metaphors that did contain explicit keywords from design metaphors to be a good reflection of their own experiences. We find that this is true for only one platform and one design metaphor in particular---ChatGPT's \plainquote{\textbf{assistant}} metaphor. Every user metaphor for ChatGPT that contained the \plainquote{\textbf{assistant}} keyword was rated in the top quartile of user metaphors for the platform. Furthermore, the keyword directly translated to multiple user metaphors, seven in total, as seen in Figure~\ref{fig:results-matches}. 

In addition, we find that more than just the explicit keyword matters, but also the associations that users communicate between the metaphor and the platform. Specifically, we see instances where references of the same keyword in user metaphors are rated differently. For example, the explicit keyword \plainquote{\textbf{community}} is seen across three user metaphors for YouTube. However, \plainquote{YouTube is a \textbf{community} because there's lots of people to connect with}~($\mu=3.31$) and \plainquote{YouTube is a \textbf{community} hub because it brings people together}~($\mu=3.19$) are rated as being \textit{less} reflective of users' experiences than \plainquote{YouTube is a \textbf{community} because I can find creators (and followers) for nearly every topic I'm interested in}~($\mu=4.04$). We also see the \textbf{community} keyword across designer and user metaphors for Twitter. However, the user metaphor \plainquote{Twitter is a more diverse thought \textbf{community} because the censorship has stopped}~($\mu=2.16$) is rated as being less reflective than both \plainquote{Twitter is a \textbf{community} because people can come together and share ideas}~($\mu=3.47$) and \plainquote{Twitter is \textbf{community} engagement because it can bring others together}~($\mu=3.29$).

\subsubsection{User Metaphors Related to Keywords from Design Metaphors}
\label{results:related-match}
Given that few user metaphors contained explicit keywords from design metaphors, we also identified user metaphors that were similar to design metaphors but may not have contained the explicit keywords. We found 8 user metaphors that were similar to design metaphors for ChatGPT, 21 that were similar for Twitter, and 27 that were similar for YouTube. Only three of our design metaphors had no similar user metaphors within our dataset - \plainquote{Twitter is a \textbf{telegraph}}, \plainquote{ChatGPT is a \textbf{sibling} model}, and \plainquote{ChatGPT is the latest \textbf{step}}. We also explored the ratings of user metaphors related to keywords from design metaphors, finding some that are in the top quartile of rated metaphors for their respective platforms.

For example, the highest rated user metaphor with explicit keywords for YouTube was \plainquote{YouTube is a \textbf{community} because I can find creators (and followers) for nearly every topic I'm interested in}~($\mu=4.04$). However, by also searching for related user metaphors, we find that the metaphor \plainquote{YouTube is entertainment because it allows me to escape to a different place and experience new things} is similar to the design metaphor \plainquote{YouTube is an entertainment \textbf{destination}} and rated slightly higher on average~($\mu=4.24$) than the community user metaphor surfaced from only matching explicit keywords. In fact, our search for user metaphors related to design metaphors surfaced the highest rated metaphor for Twitter---\plainquote{Twitter is a chaotic town square because everyone can communicate out their ideas but it is difficult to tell fact from fiction}~~($\mu=3.96$). We argue that this metaphor is related to the design metaphor \plainquote{Twitter is a \textbf{community}}. Finally for ChatGPT, we see that similar metaphors are roughly just as reflective as metaphors with explicit keywords, adding new vocabulary for what might be considered an \plainquote{\textbf{assistant}} (e.g., \plainquote{ChatGPT is a helpful adult}). 

Despite this expanded search, we only surfaced the highest ranked user metaphor for one of our three platforms, Twitter (via \plainquote{Twitter is a chaotic town square}). Moreover, none of the user metaphors that referenced explicit keywords from the design metaphors captured the highest rated user metaphor for any platform. Figure~\ref{fig:results-matches} reflects these findings and also shows a great deal of spread across the ratings of user metaphors that are related to design metaphors, much like the spread across the ratings of user metaphors with explicit keywords from design metaphors. This suggests that design metaphors do not always resonate in the few instances they translate.

\subsection{Reflective Metaphors for Users' Experiences}
\figcodes
To further understand the differences between design and user metaphors, we qualitatively grouped both types of metaphors to surface differences in their underlying experiences. We outline the groups for each platform in Figure~\ref{fig:results-codes}. As indicated by our quantitative results in Section~\ref{results:match}, design metaphors often communicate different intended experiences than the experiences reflected in user metaphors. For instance, there is some overlap between the focus on communicating across Twitter's design metaphors and the experience of being in a conversational room evoked by nine of Twitter's user metaphors. However, user metaphors also presented experiences around self-expression, challenging viewpoints, and toxicity that are not communicated in Twitter's design metaphors. Beyond diverging from the experiences communicated by design metaphors, user metaphors also communicate user experiences that converge and diverge from each other. For example, 15 of the 41 highest rated user metaphors for ChatGPT focused on the abundance of knowledge across users' interactions with ChatGPT. The most common metaphor for this experience was a \plainquote{library} (e.g.: \plainquote{ChatGPT is a library because it provides a vast array of info and knowledge to explore}). Yet, \plainquote{an encyclopedia} and \plainquote{a cyber brain} were also used to characterize this experience. In contrast, the experience of endless exploration, described by some of the user metaphors for YouTube, contained both positive and negative perspectives on this experience. For instance, the metaphors \plainquote{YouTube is a treasure chest because you can endlessly explore} and \plainquote{YouTube is a rabbit hole because you can get lost in the algorithm very easy} highlight the benefits and challenges of the experience. We see a similar example of this for Twitter in the which the experience of self-expression was reflected among 9 of 45 metaphors with the highest ratings across different sentiments: \plainquote{Twitter is a free space because anyone can say anything}~($\mu=3.54$) and \plainquote{Twitter is the wild west because people feel they can say anything online}~($\mu=3.84$).

\subsection{Different Experiences Across Different Groups}
These diverging experiences and sentiments may indicate differences in how different user communities rate metaphors. As such, we identify metaphors with different means across racial, gender, and age groups using ANOVA and identify the specific group differences for these metaphors by running unpaired t-tests across group pairs. Across user metaphors, we find 54 metaphors that users from different racial groups rate higher than users from other groups, 18 metaphors that men and women rate differently, and 8 metaphors that reflect experiences of various age groups differently. Notably, we find a few metaphors with average ratings in the top quartile that have significantly different ratings across communities. We highlight these metaphors in Table~\ref{tab:demo-differences}, and include the full list of differences across metaphors in the supplementary materials.

\renewcommand{\arraystretch}{1.35}
\begin{table}[!ht]
    \caption{User metaphors with significant rating differences across groups. We see that even some metaphors with average ratings in the top quartile (indicated with a dark yellow triangle~\topQuartile) exhibit differences in their reflectivity for users from different communities. We indicate significance using the following: \textbf{*}$p_{corrected}<0.05$; \textbf{**}$p_{corrected}<0.01$. Please note that negative differences indicate that group B rated the metaphor higher on average.}
    \label{tab:demo-differences}
    \centering
    \small
    \begin{tabular}{p{0.45\linewidth}ccc}   
    \toprule
    \textbf{User Metaphor} & \textbf{A} & \textbf{B} & \textbf{$\mu_A-\mu_B$}\\ \midrule
    ChatGPT is \textbf{a compass} because it helps guide people through information & Black & Asian & 1.09\textbf{*}\\
    ChatGPT is \textbf{a compass} because it helps guide people through information & Black & White & 1.49\textbf{*}\\
    ChatGPT is \textbf{an encyclopedia} because it provides answers to my questions~\topQuartile &  Black & Hisp./Latino & 1.64\textbf{**}\\
    Twitter is \textbf{a maze} because there are so many false paths before getting the information you want~\topQuartile & Asian & White & 1.19\textbf{*}\\
    Twitter is \textbf{a town hall} because people can openly express their thoughts~\topQuartile & Asian & Black & -1.39\textbf{*}\\
    YouTube is \textbf{a training ground} because I can find instructions for nearly anything I want to do~\topQuartile & Asian & Hisp./Latino & -1.1\textbf{*}\\
    YouTube is \textbf{a community hub} because it brings people together & Black & Asian & 1.54\textbf{**} \\ \midrule
    ChatGPT is \textbf{a teacher} because it shows you other ways to say things~\topQuartile & Women & Men & -0.81\textbf{*}\\
    Twitter is \textbf{a diary} because many share their inner most thoughts~\topQuartile & Women & Men & 0.75\textbf{*}\\
    Twitter is \textbf{a rabbit hole} because there are many different threads of information to go down~\topQuartile & Women & Men & 0.66\textbf{*}\\
    Twitter is \textbf{a canvas of thoughts} because you can express whatever comes into you mind in writing~\topQuartile & Women & Men & 0.75\textbf{*}\\ \midrule
    ChatGPT is \textbf{a teacher} because it provides answers to questions & 18-24 & 45-54 & -2.43\textbf{*}\\
    ChatGPT is \textbf{a forest fire} because it keeps destroying the planet and it's very hard to stop once it's started going & 18-24 & 35-44 & 1.4\textbf{*}\\
    YouTube is \textbf{a library} because you can learn almost anything there~\topQuartile & 25-34 & 18-24 & 1.04\textbf{*}\\
    YouTube is \textbf{a library} because you can learn almost anything there~\topQuartile & 25-34 & 35-44 & -1.17\textbf{*}\\ \bottomrule
    \end{tabular}
\end{table}

\subsubsection{Differences Across Racial Groups} Across all three platforms, we find 54 metaphors that are more reflective of the experiences of one racial group than those of another. For example, Black users rated several knowledge-based metaphors as being more reflective of their experiences than their White counterparts did. There was a significant difference between how Black users ($\mu=3.92$) and White users ($\mu=2.43$) rated \plainquote{ChatGPT is a compass because it helps guide people through information}~($t=3.99,~p_{corrected}=0.01$). There were also significant differences across metaphors such as \plainquote{ChatGPT is a wise man because it holds knowledge we find valuable}~($t=3.66,~p_{corrected}=0.01$), \plainquote{ChatGPT is a second brain because it helps you think outside your own box}~($t=4.12,~p_{corrected}=0.007$), and \plainquote{ChatGPT is a guardian because it always has an answer and is willing to help}~($t=4.65,~p_{corrected}=0.003$). In fact, across all three platforms, many of the differences in rating across racial groups stemmed from Black users rating metaphors as more reflective of their results than users from other communities.

\subsubsection{Differences Across Gender Groups} While fewer in number, we found significant differences between how men and women rated 18 metaphors across ChatGPT and Twitter. For instance, there was a significant difference between how women~($\mu=4.00$) and men~($\mu=3.25$) rated \plainquote{Twitter is a canvas of thoughts because you can express whatever comes into you mind in writing}~($t=-2.264,~p_{corrected}=0.03$). Conversely, we found a significant difference between how men~($\mu=3.44$) and women~($\mu=2.69$) rated \plainquote{Twitter is radio because you can tune into new talking points}~($t=-2.16,~p_{corrected}=0.046$). Such differences outline different use cases across the two groups. However, we found no significant differences between how men and women rate YouTube user metaphors.

\subsubsection{Differences Across Age Groups} We found significant differences across how different age groups rated 8 metaphors across all three platforms. Notably, there are differences across several age groups for the same metaphors. For example, there is a significant difference between how users between the ages of 18 and 24~($\mu=4.1$) and users between the ages of 25 and 34~($\mu=3.06$) rated \plainquote{YouTube is a library because you can learn almost anything there.}~($t=3.037,~p_{corrected}=0.0058$). This trend continues, as 25-34 year-olds also rate the same metaphor significantly lower than 35-44 year olds~($t=-3.11,~p_{corrected}=0.04774$).

\section{Discussion}
In this work, we aim to offer a new way of assessing metaphor quality by considering how design and user metaphors for platforms compare to one another. We find that many of the design metaphors used to convey the intended experiences of our three platforms rarely translate directly to the experiences of users. Even when a design metaphor is captured in users' metaphors, user ratings indicate that the system may not reflect or only partially reflects the design metaphor. For instance, while the simplified design metaphor \plainquote{YouTube is a \textbf{community}} is directly reflected in some user metaphors, users' experiences better reflect the aspect of finding interesting people (\plainquote{YouTube is a community because I can find creators (and followers) for nearly every topic I'm interested in}) than the aspect of actually connecting and building relationships (\plainquote{YouTube is a community because there's lots of people to connect with}). Overall, we found only one design metaphor that 1) translated directly to multiple user metaphors and 2) was consistently rated as being highly reflective of users' experiences:~\plainquote{ChatGPT is an \textbf{assistant}}. Moreover, only one platform in our study - Twitter - had a user metaphor that was both related to the platform’s design metaphor and the highest ranked metaphor - namely the metaphor \plainquote{Twitter is a chaotic town square}. However, none of the user metaphors that referenced explicit keywords from the design metaphors captured the highest rated user metaphor for any platform.

With this in mind, we qualitatively coded user and design metaphors to understand how the underlying experiences differed across both sets of metaphors. We find that user metaphors reflect a number of experiences beyond those embedded in design metaphors. For example, user metaphors for Twitter detailed experiences around toxicity and navigating information and misinformation. User metaphors for ChatGPT denoted interests in gaining knowledge and not just completing tasks. Learning, content diversity, and exploration were notable themes among the top rated user metaphors for YouTube--themes that go beyond YouTube's design metaphors. We argue that these more varied experiences and metaphors better reflect what users find valuable.

Furthermore, our findings show that metaphor ratings are diverse across groups. For one, even when users share the same experience, they may have different sentiments about the experience or only connect with part of an experience. Moreover, communities have different and distinct experiences on the same platform, even with metaphors that have high average ratings. This echoes prior work that questions the prevalence of the unmarked user in design and challenges practitioners and researchers alike to consider the different experiences of people from diverse racial, gender, and age backgrounds~\cite{costanza-chock_design_2020, bardzell_feminist_2010, ogbonnaya-ogburu_critical_2020}. Together, these findings add to the existing challenges that designers face when identifying metaphors that convey and communicate their intended experience for a system.

However, we argue that this work also creates opportunities to support design practitioners in communicating their intended experiences to a wide array of users. We detail these opportunities below.

\subsection{Opportunity 1: Metaphor Comparisons for Understanding and Refining User Experience}
While metaphors in HCI and design have primarily been relegated as a tool for designers, this work extends prior work to add new ways for user-generated metaphors to support design processes~\cite{french_whats_2017,desai_metaphors_2023,cheng_tools_2025}. In particular, our work shows that users can navigate the nebulousness of user experience~\cite{alves_state_2014,bargas-avila_old_2011,law_understanding_2009,robinson_past_2018} and develop and rate multiple metaphors to articulate their experiences with systems. Metaphors may be an ideal tool for such feedback, as they enable people to understand the abstractness of user experience. Furthermore, these methods enable users to communicate their experiences in their own words and indicate the saliency of their experiences. We also see potential in utilizing user-generated metaphors to shape more socially-responsible and community-centered systems, by comparing and contrasting user metaphors across different communities. Notably, our method incorporates lightweight, mixed-method approaches that make collecting such feedback accessible for options for designers that need to balance constrained resources, diverse data, and digestible insights~\cite{kang_challenges_2024}.

Furthermore, comparing metaphors may enable design teams and researchers to interrogate what aspects of a metaphor are (not) relayed to the user. Prior work focused on usability has described the ``missing'' parts of metaphors---or the aspects of a metaphor's functionality that do not map to a system's functionality---as conceptual baggage~\cite{anderson_minimising_1994}. Our findings suggest that there are similar, ``missing'' parts when it comes to metaphors that reflect user experiences. Specifically, our method surfaced several instances where users noted different saliencies across metaphors that were similar but differed in their explanations and associations. These differences may support designers in revising features to strengthen aspects of their underlying design metaphor. Surfacing these differences also combats the idea of neutral design artifacts, as users often describe the mixed experiences of a metaphor, such as the utility but ``unnaturalness'' of users' \textit{robot} metaphors for ChatGPT. In particular, comparing design and user metaphors may further ongoing design and research around metaphors in AI~\cite{cheng_tools_2025, dove_monsters_2020, gilardi_we_2024, yang_ai_2025, kajava_language_2023} by surfacing which aspects of similar human-related metaphors are highly salient with users and how those aspects may be altered or magnified through design.

\subsection{Opportunity 2: Supporting Iterative Metaphor Generation}
In addition to supporting user input around metaphors and user experience, our work also presents the opportunity for more iterative metaphor generation and refinement. While we focus on evaluating user experience via user-generated metaphors, these findings can spur refinement of existing design metaphors and/or entirely new ideas for user experiences. For example, if users communicate appreciation for the diary-like experience of platform that was shaped to be an information network, how might designers alter underlying system metaphors to further support users? Could the metaphors be combined into a composite metaphor~\cite{hsu_approach_2007} or could designers generate new metaphors that combine similar attributes? Designers could further evaluate a resulting prototype from these revised metaphors and ask a different round of participants to re-evaluate this updated experience. Additionally---how do changes in design shift the metaphors that users see in a platform? For example, Twitter, Bluesky, Threads, and Mastodon share many design features---but not all~\cite{zhang_form-_2024}. Are these design differences associated with differences in user metaphors? If Threads changed its design, how much would that change the metaphors that users see in the platform? In this way, our work lays the foundation for more iterative metaphor generation by providing a template for user metaphor evaluation.

\subsection{Opportunity 3: Longitudinal shifts in metaphors}
Prior work has suggested that metaphor elicitation via sentence-based completion methods can help to monitor user perspectives over time~\cite{cheng_tools_2025}. Our methods and findings suggest that we may also be able to explore longitudinal shifts across design metaphors and their different user metaphors. As platforms evolve over time, so do design metaphors and users' experiences with these platforms. For instance, within a year or two of launch, Twitter was centered around casual conversations among friends, families, and strangers, likening its user experience to that of a communication service, a \plainquote{telegraph system of Web 2.0}, and a \plainquote{global community of friends and strangers}. However, as the platform grew, Twitter's metaphors shifted to that of a \plainquote{real-time information network} focused on both critical updates and users' broader interests.

Even for ChatGPT, a platform with a much shorter history than Twitter or YouTube, we are able to see changes in how platforms describe their experiences. The about page during the research release of ChatGPT presented the platform as a \plainquote{sibling model} to InstructGPT, drawing on language that would be familiar to an academic audience that regularly employs language like \textit{parent}, \textit{child}, \textit{master} (problematically), and even \textit{forks} to articulate relationships between objects. In particular, this \plainquote{sibling} metaphor alludes to the shared training methods for both models, but the diverging data collection - as GPT-3's training data was more focused on dialogue~\cite{openai_introducing_2022} while InstructGPT's training resulted in better instruction-following~\cite{ouyang_training_2022, openai_aligning_2022}. Furthermore, the platform indicated its broader plan for iteration by exemplifying ChatGPT as a \plainquote{step} toward more powerful systems in the future~\cite{openai_introducing_2022}. However, as the platform was adopted by a more general and industry-focused audience, we not only see a change in the interface design of the about page, but also in the design metaphor - which conveys ChatGPT as a \plainquote{superassistant}.

Future work can adopt these methods to surface patterns of change in design and user metaphors over time. For example, time series analyses might help us tease out causality: do designers' metaphors come earlier than users' metaphors, or do users generate the metaphors first and then platforms adopt them and amplify them? Likewise, exploring shifts in design and user metaphors can help to contextualize shifts in societal narratives and conversations around artificial intelligence~\cite{gilardi_we_2024}, toxicity, privacy, and social media.

\subsection{Limitations}
Our study identified gaps between the metaphors that platforms use to describe their user experience, and the metaphors that users use to describe their experiences with platforms. To maintain scope, we restricted our study to three major platforms---future work can expand the study to additional platforms to identify even larger-scale patterns. Likewise, our user metaphors were gathered at a single time point. We advocate for future work that gathers time-series data to help understand how metaphor ratings change over time, and whether users grow more in sync, or less in sync, with platform metaphors over time. 

To map design metaphors onto user metaphors, our study relied on two independent annotators rating metaphor similarity and resolving differences through discussion. While this is a common method, metaphors can reference cultural concepts that are familiar to some but unknown to others---so development of future methods could help researchers classify metaphor similarity with increasing accuracy.

Finally, our analysis recruited participants in the United States, which focuses our analytical frame in North America. Metaphor differences are likely to be accentuated in international cultural contexts. Cross-national metaphor elicitation surveys could help make our results more globally robust---and measure whether, as we suspect, platform metaphors resonate most strongly with a North American audience.
\section{Conclusion}
In this paper, we examine whether design metaphors for major online platforms resonate with users' experiences of the platform. To do so, we performed archival web research to identify metaphors that platform designers have used to describe the intended experience of their systems, and compared these results to a metaphor elicitation survey of platform users. We observed that many design metaphors do not translate directly to users' experiences. Even when user mention the same metaphors as those projected by platforms, users' experiences may only align with certain aspects of the design metaphor. We also observed metaphors from users that captured experiences beyond those reflected in design metaphors, including mixed and negative experiences with the platforms. Through this work, we look to show how comparing design and user metaphors creates common ground for users, communities, and platforms to find improved communication around the strengths and weaknesses of a design.

\begin{acks}
We appreciate our anonymous reviewers for their support in refining this work. We thank Jordan Troutman and Poonam Shah for their support annotating data. Furthermore, we are grateful for members of the Stanford HCI Group - especially Jordan Troutman, Nava Haghighi, Michelle Lam, Yutong Zhang, and Jane E - for their valuable feedback on various versions of this paper. This work was supported in part by the Stanford Institute for Human-Centered Artificial Intelligence, the Hasso Plattner Foundation and by NSF grant CCF-1918940. Beleicia Bullock was supported by the Stanford Center of Philanthropy and Civil Society's Ph.D. Fellowship and the Center for Comparative Studies in Race and Ethnicity's Technology and Racial Equity Graduate Fellowship.
\end{acks}

\bibliographystyle{ACM-Reference-Format}
\bibliography{references}

\appendix

\end{document}